\documentclass[amssymb,twocolumn,aps,prstper,superscriptaddress,english,nofootinbib]{revtex4}
\usepackage{graphics,color,graphicx,amsmath}
\usepackage{subcaption}
\usepackage{natbib}
\usepackage{verbatim}
\usepackage{graphicx}
\usepackage[above,below]{placeins}
\usepackage{float}
\usepackage{tabularx}
\usepackage{blindtext}
\usepackage{url}
\usepackage{rotating}
\usepackage{dcolumn}

\usepackage{hyperref}
\hypersetup{
    colorlinks=blue,
    linkcolor=blue,
    filecolor=blue,      
    urlcolor=blue
    }

\begin{document}

\title{Bias in physics peer recognition does not explain gaps in perceived recognition}

%\title{Bias in physics peer recognition does not explain gaps in perceived recognition}
%\title{Testing models of physics peer recognition and perceived recognition}

% Direct comparison of undergraduate science students' actual and perceived peer recognition; For undergraduate science students, men and women internalize their peer recognition differently

\author{Meagan Sundstrom}
\affiliation{Laboratory of Atomic and Solid State Physics, Cornell University, Ithaca, New York 14853, USA}
\affiliation{Department of Physics, Drexel University, Philadelphia, Pennsylvania 19104, USA}

\author{N. G. Holmes~\footnote[1]{Corresponding author. ngholmes@cornell.edu}}
\affiliation{Laboratory of Atomic and Solid State Physics, Cornell University, Ithaca, New York 14853, USA}

\begin{abstract}
Gaining recognition as a physics person by peers is an important contributor to undergraduate students' physics identity and their success in physics courses. Previous research has separately demonstrated that women perceive less recognition from peers than men in their physics courses (perceived recognition) and that women receive fewer nominations from their peers as strong in their physics course than men (received recognition). The relationship between perceived and received peer recognition for men and women, however, is not well understood. Here we test three plausible models for this relationship. We conduct a large-scale, quantitative study of over 1,700 students enrolled in introductory physics courses at eight institutions in the United States. We directly compare student gender, perceived recognition, and received recognition, controlling for other student demographics and course-level variability. Results show with high precision that, for students receiving the same amount of recognition, and having the same race or ethnicity, academic year, and major, women report significantly lower perceived recognition than men. These findings offer important implications for testable instructional interventions. 
\end{abstract}

\maketitle

The under-representation of women and people of color in undergraduate physics courses and, subsequently, in physics careers is well-documented~\cite{clark2005,sax2016,porter2019women,merner2015african}. Broadening the population of physicists would give voice to new and innovative ideas that are critical to scientific advancements~\cite{tilghman2021concrete,powell2018these,cochran2024racial}. While many structural barriers to entry and retention in physics exist for women in particular, one significant challenge is that women may both \textit{perceive}~\cite{lock2013physics,kalender2019female,kalender2019gendered,hazari2013science,bottomley2022relationship} and \textit{receive}~\cite{grunspan2016,salehi2019,bloodhart2020,sundstrom2022perceptions,sundstromWhoWhat} less recognition from their physics peers about their abilities as physicists than men. Perceived recognition from peers directly correlates with students' development of their physics identity (i.e., the extent to which they believe they are a ``physics person"~\cite{carlone2007understanding}) and their persistence in physics courses and careers~\cite{hazari2010connecting,kalender2019female,kalender2019gendered,lock2013physics,cech2011professional}.

Existing research in physics education, however, has only examined gender differences in students' perceived and received peer recognition separately~\cite{grunspan2016,salehi2019,bloodhart2020,sundstrom2022perceptions,sundstromWhoWhat,lock2013physics,kalender2019female,kalender2019gendered,hazari2013science,bottomley2022relationship}. Understanding the nuance of how students internalize the peer recognition they actually receive is crucial for identifying testable instructional interventions aimed at mitigating gaps in perceived recognition between men and women. In this study, therefore, we directly compare the extent to which students feel recognized by their peers (\textit{perceived recognition}) to the number of nominations they receive from their peers for being strong in their physics course (which we assume is a proxy for \textit{received recognition}), as indicated on surveys. We operate under an equity of parity (or equity of outcomes) model where the instructional goal is for men and women to report comparable levels of perceived recognition, such that men and women have similar opportunities for identity development, success in physics courses, and retention in physics~\cite{RodriguezEquity,BurkholderEquity}.

\begin{figure*}[t]
    \centering
    \includegraphics[width=6.7in]{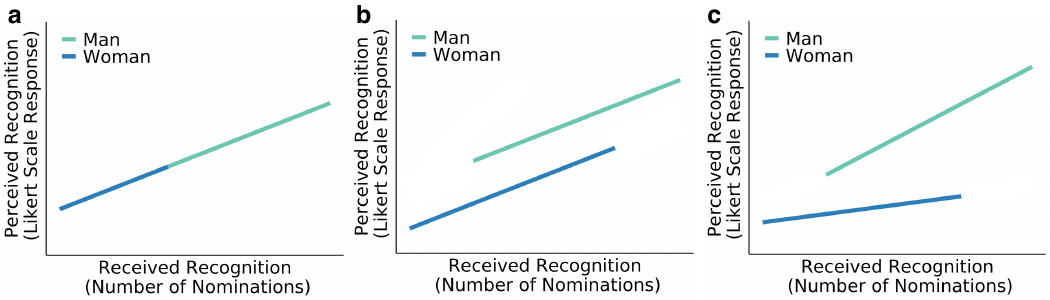}
    \caption{Toy quantitative relationships between student gender, perceived recognition, and received recognition.}
    \label{fig:hypotheses}
\end{figure*}

We directly test three possible quantitative relationships between perceived and received peer recognition for men and women (Fig.~\ref{fig:hypotheses}). The first relationship proposes that both men and women internalize their peer recognition similarly. Under this relationship, men only perceive more peer recognition than women~\cite{lock2013physics,kalender2019female,kalender2019gendered,hazari2013science,bottomley2022relationship} because they receive more peer recognition than women (Fig.~\ref{fig:hypotheses}a), as suggested by prior work~\cite{grunspan2016,bloodhart2020,sundstrom2022perceptions,sundstromWhoWhat}. If this relationship best describes our data \textit{and} our assumption that student nominations of strong peers made on a survey are an appropriate proxy for received recognition is true, then instructional intervention experiments should target received recognition. Visually, an intervention aimed at shifting the horizontal ranges of men's and women's lines in Fig.~\ref{fig:hypotheses}a would be more effective than one aimed at vertically shifting men's and women's lines to create overlapping lines for men and women (in line with the equity of parity goal). To target received recognition, instructors may try to refine students' criteria through which they recognize their physics peers or diversify students' peer networks in order to alleviate any gender biases in students' nominations of strong peers. 

The second relationship, in contrast, is that men and women internalize their peer recognition differently. Under this relationship, men perceive more peer recognition than women~\cite{lock2013physics,kalender2019female,kalender2019gendered,hazari2013science,bottomley2022relationship} because they have higher perceptions of their peer recognition than women receiving the same amount of recognition (though men may also receive more peer recognition
than women; Fig.~\ref{fig:hypotheses}b). If this relationship best describes our data (and our assumption that student nominations of strong peers made on a survey are an appropriate proxy for received recognition is true), instructional intervention experiments should target perceived recognition to ensure that all students appropriately internalize recognition from their physics peers. Visually, an effective intervention would aim to vertically shift men's and women's lines in Fig.~\ref{fig:hypotheses}b to be overlapping (a supplemental intervention aimed at shifting the horizontal ranges of men's and women's lines in Fig.~\ref{fig:hypotheses}b may also be appropriate if there is also a gender bias in received recognition). To target perceived recognition, instructors may help students identify different forms of recognition from others and/or give students time to reflect about the recognition they have received from others.

The third relationship is that men and women internalize their peer recognition at different rates. Under this relationship, the difference between men's and women's perceived recognition is different at different amounts of received recognition (and men may also receive more peer recognition than women; Fig.~\ref{fig:hypotheses}c). If this relationship best describes our data (and our assumption that student nominations of strong peers made on a survey are an appropriate proxy for received recognition is true), the appropriate instructional intervention experiment would depend on the particular relationship. An intervention should target perceived recognition (i.e., vertically shifting men's and women's lines in Fig.~\ref{fig:hypotheses}c), for example, if the difference between men's and women’s perceived recognition increases for higher amounts of received recognition, as shown in Fig.~\ref{fig:hypotheses}c. On the other hand, an intervention targeting received recognition (i.e., horizontally shifting men's and women's lines in Fig.~\ref{fig:hypotheses}c) might be more effective if the difference between men's and women’s perceived recognition decreases for higher amounts of received recognition.

To test these relationships, we evaluate data collected from introductory physics students via an online survey, using existing measures of perceived~\cite{hazari2010connecting,lock2013physics,kalender2019gendered,hazari2013science,bottomley2022relationship} and received~\cite{grunspan2016,salehi2019,bloodhart2020,sundstrom2022perceptions,sundstromWhoWhat} peer recognition. The perceived recognition item asks students the extent to which they agree with the statement ``My peers see me as a physics person'' on a 5-point Likert scale ranging from ``Strongly disagree'' to ``Strongly agree.'' The received recognition item asks students to nominate peers who they feel are ``particularly strong" in their physics course, through which we determine the number of nominations received by each student. We ask these questions separately for the instructional contexts of laboratory (lab) and lecture, when applicable, because prior work has demonstrated that peer recognition forms differently in these two instructional contexts--even among the same set of students--given their different course structure and learning objectives~\cite{sundstrom2022perceptions}. 

We perform our analysis with 1,721 unique students across 27 introductory physics contexts (hereon referred to as individual ``courses"): 15 instructional lecture courses (containing 1,059 total students; 559 men, 480 women, and 20 students of additional gender identities) and 12 instructional lab courses (containing 1,683 total students; 794 men, 858 women, and 31 students of additional gender identities; see Table~\ref{tab:demographics}). The courses come from eight institutions in the United States. These institutions are diverse across multiple dimensions, including minority-serving status, public and private institutions, and geographic regions (see Methods). All institutions, however, are PhD-granting and in the United States. %While the courses are all at the introductory level, students come from all four academic years, a range of science and engineering majors, and identify with a range of races or ethnicities (see Table~\ref{tab:demographics}). 

\begin{figure*}[hp!]
    \centering
    \includegraphics[width=6.2in]{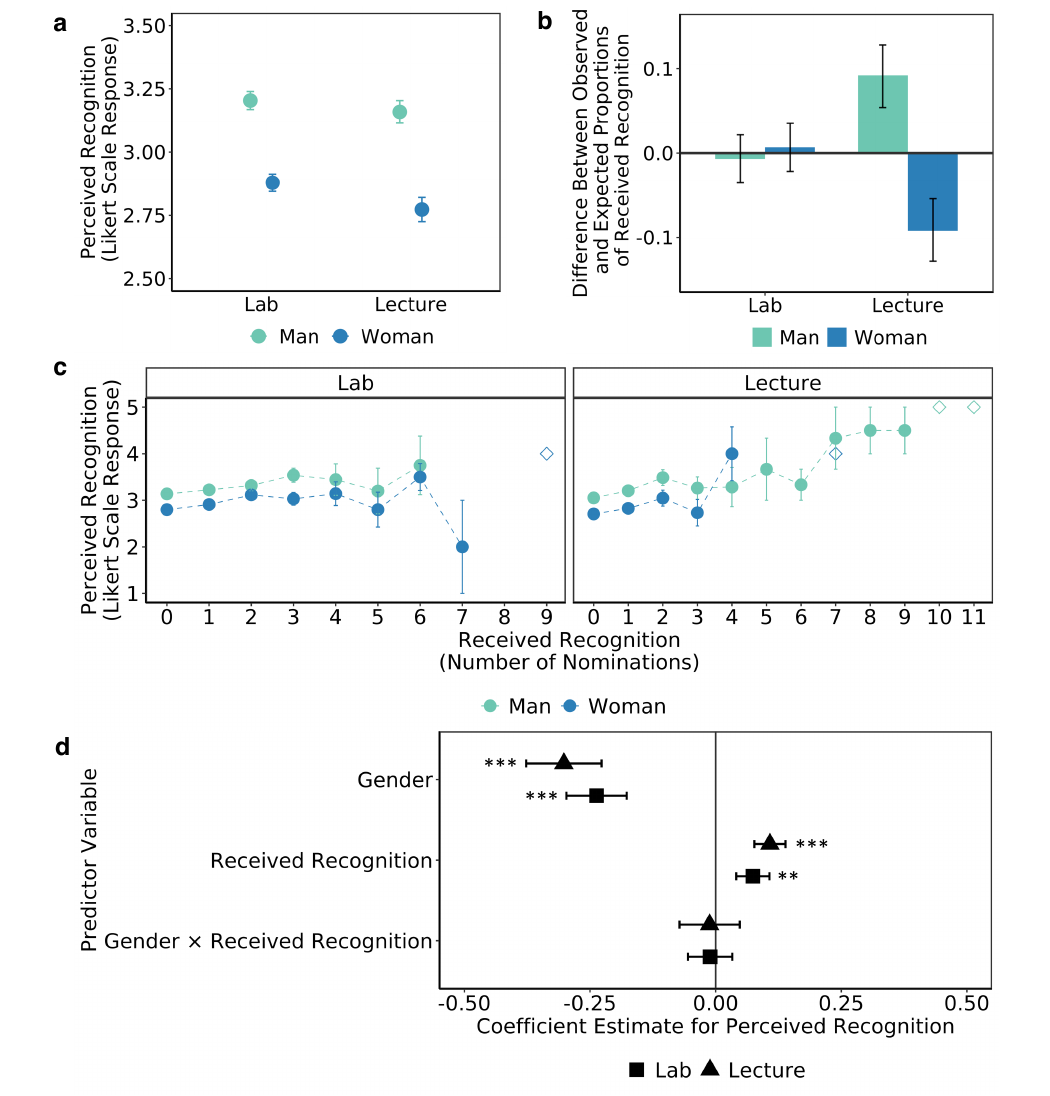}
    \caption{Study results. (a) Dot plots of mean perceived recognition by student gender and instructional context. Likert-scale agreement was converted to a 5-point ordinal scale with 1 representing ``Strongly disagree" and 5 representing ``Strongly agree." Error bars represent the standard uncertainty in the mean.  (b) Bar charts comparing the observed proportion of recognition received by men or women, measured as the proportion of total nominations received by men or women, to the expected unbiased proportion of received recognition for men or women, calculated as the proportions of men and women, split by instructional context. Positive (negative) values indicate that the observed proportion is higher (lower) than the expected unbiased proportion. Error bars represent 95\% confidence intervals on the  observed proportions.  (c) Line plots comparing mean perceived recognition at each level of received recognition, split by student gender and instructional context. Unfilled diamonds represent individual students. Error bars represent the standard uncertainty in the mean. (d) Results of linear mixed models for three of the predictor variables (see full model results in Table~\ref{tab:coefficients}). The reference group for the gender term is men, thus the coefficients shown are for women relative to men. Error bars represent standard errors of the coefficient estimates and asterisks indicate statistical significance ($^{**} p < $0.01; $^{***} p < $0.001). }
    \label{fig:results}
\end{figure*}

All 27 courses had a survey response rate above 70\% of enrolled students. We exclude data collected from an additional 47 introductory physics courses (34 instructional lecture courses and 13 instructional lab courses, all from the same set of eight institutions) with a response rate below 70\% because response rates below this threshold render our measure of received recognition too sensitive to missing data~\cite{smith2013structural}. We report the results for students who self-identify as a man or a woman because there were not enough students of additional gender identities to make meaningful statistical claims about them. We do, however, include these students in our statistical models and discuss their preliminary results in the Supplementary Information. 

%prior work predicts recognition differences between these two groups--there are not yet any established expectations of recognition differences between other gender identities.

We note that we examine students pursuing a variety of majors within their physics courses. Physics identity, however, is associated with students' success in their physics courses, and thus is valuable for any student who is required to take physics. Moreover, physics identity has been shown to be important for the career choices of engineering students, who comprise a large fraction of our sample~\cite{godwin2016identity}. Non-physics majors' experiences in physics courses can also shape their other disciplinary identities and intentions to pursue their chosen field of study~\cite{patrick2016review}. 

\section*{RESULTS}

\begin{figure*}[hbt!]
    \centering
    \includegraphics[width=6in]{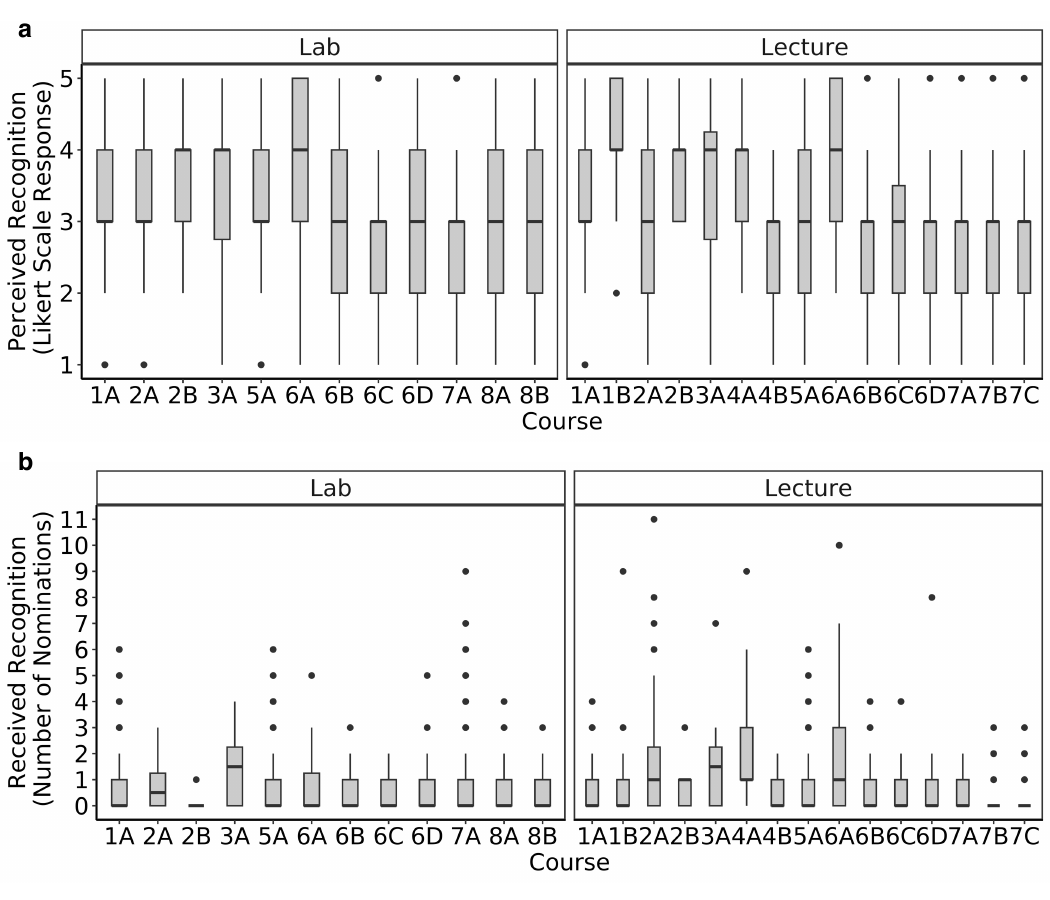}
    \caption{Distributions of (a) perceived recognition and (b) received recognition split by individual course within each instructional context. The grey boxes indicate interquartile range with the bold lines representing the medians. Whiskers denote $1.5\times$ the interquartile range and additional points indicate outliers. %There is no systematic variation of medians, interquartile ranges, and outliers of the distributions despite the wide range of class sizes, thus we aggregated the data from all courses within each instructional context.
    }
    \label{fig:disaggregated}
\end{figure*}

Similar to existing research~\cite{lock2013physics,kalender2019female,kalender2019gendered,hazari2013science,bottomley2022relationship}, Student’s \textit{t}-tests show that, on aggregate, men report significantly higher perceptions of peer recognition than women in both the lab (\textit{t} = 6.60, df [degrees of freedom] = 1639.6, \textit{p} $<$ 0.001, Cohen’s $d$ (effect size) = 0.33) and lecture (\textit{t} = 5.92, df = 1015.9, \textit{p} $<$ 0.001, Cohen’s $d$ = 0.37) contexts in this study (Fig.~\ref{fig:results}a; see distributions in Figs.~\ref{fig:disaggregated}a and~\ref{fig:aggregateperceived}). 

Contrary to some prior work~\cite{sundstrom2022perceptions}, we find that men and women receive proportional recognition in the lab context: the fractions of nominations made to men and women do not significantly differ from the fractions of men and women in the sample. Visually, the error bars of the proportions in lab in Fig.~\ref{fig:results}b overlap with zero (see full distributions  of received recognition in Figs.~\ref{fig:disaggregated}b and~\ref{fig:aggregateactual}). The same analysis shows that, on aggregate, men (women) receive disproportionately more (less) recognition than expected in the lecture context: the fraction of nominations received by men is much larger than the fraction of men in the sample, with a small effect size. Visually, the error bars of the proportions in lecture in Fig.~\ref{fig:results}b do not overlap with zero.

We use linear mixed models to directly test the three possible relationships shown in Fig.~\ref{fig:hypotheses}, with perceived recognition as the dependent variable. The models include predictor variables for gender, received recognition, and the interaction between gender and received recognition, as well as students’ academic year, academic major, and race or ethnicity (see full model results in Table~\ref{tab:coefficients}). The models also include course as a random intercept. 

We find that the second relationship (Fig.~\ref{fig:hypotheses}b) best describes our data set for both instructional contexts (Figs.~\ref{fig:results}c and~\ref{fig:results}d).  That is, for students receiving the same amount of peer recognition (and having the same academic year, academic major, and race or ethnicity), women report significantly lower perceived recognition than men (lab: \textit{b} = --0.24 [95\% confidence interval: --0.35, --0.12], SE = 0.06, \textit{t} = --3.92, \textit{p} $<$ 0.001; lecture: \textit{b} = --0.30 [--0.45, --0.15], SE = 0.08, \textit{t} = --4.03, \textit{p} $<$ 0.001). Received recognition also separately predicts perceived recognition: students receiving more recognition (and having the same gender, academic year, academic major, and race or ethnicity) report significantly higher perceptions of their recognition in both contexts (lab: \textit{b} = 0.07 [0.01, 0.14], SE = 0.03, \textit{t} = 2.25, \textit{p} $<$ 0.05; lecture: \textit{b} = 0.11 [0.05, 0.17], SE = 0.03, \textit{t} = 3.52, \textit{p} $<$ 0.001). 

Comparing the sizes of these coefficient estimates, gender has a larger effect than received recognition on perceived recognition. To achieve an increase in perceived recognition equal to the average difference between men's and women’s perceived recognition, a student would need three more nominations (i.e., a three unit increase in received recognition) in either context. Additionally, the difference between men’s and women’s perceived recognition does not vary for different amounts of received recognition as indicated by the interaction term (lab: \textit{b} = --0.01 [--0.10, 0.07], SE = 0.04, \textit{t} = --0.26, \textit{p} = 0.80; lecture: \textit{b} = --0.01 [--0.13,0.10], SE = 0.06, \textit{t} = --0.21, \textit{p} = 0.84). Visually, Fig.~\ref{fig:results}c indicates these results through the parallel but systematically offset lines for men and women, similar to Fig.~\ref{fig:hypotheses}b. One caveat is that we find no aggregate gender bias in received recognition in the lab context, indicating that in this context the lines for men and women actually span a comparable horizontal range (in contrast to what is shown in Fig.~\ref{fig:hypotheses}b).

\section*{DISCUSSION}

Results indicate that, across eight different institutions, students' varying internalization of their peer recognition better explains gender differences in student perceptions of their peer recognition than gender biases in received peer recognition. We identify that the effect of gender is larger than the effect of received recognition on perceived recognition for our sample of students. For men and women receiving the same peer recognition (and having the same gender, academic year, academic major, and race or ethnicity), women, on average, report their perceived peer recognition as 0.25 points (on a 5-point Likert scale) lower than men. A woman would need to receive three additional nominations from peers (received recognition) to increase their perceived peer recognition by this amount. 

For men and women to hold comparable perceptions of their peer recognition, therefore, instructional intervention experiments must aim to vertically shift men's and women's lines in Fig.~\ref{fig:results}c to be overlapping. In both lab and lecture contexts, this suggests that intervention experiments should target shifts in perceived peer recognition, such that all physics students appropriately internalize recognition from their peers. Shifts in received recognition (that is, horizontal shifts in men's and women's lines) would not produce overlapping lines in perceived recognition.

Notably, this recommendation is in contrast to existing recommendations related to both peer and instructor recognition, which suggest instructional intervention experiments should target received recognition to mitigate perceived recognition gaps~\cite{cwik2022not,li2023recognition,sundstrom2022perceptions}. Our statistical results indicate that such an intervention may be particularly ineffective in an instructional lab context, where men and women receive proportional recognition from their peers as strong in the course material. Our results do, however, indicate that an intervention targeting received peer recognition (i.e., aiming to horizontally shift men's and women's lines in Fig.~\ref{fig:results}c to span comparable ranges) could effectively supplement an intervention targeting perceived recognition in a lecture context, where women receive significantly less peer recognition than men. That is, interventions that shift the lines in Fig.~\ref{fig:hypotheses}b vertically would result in the situation represented by Fig.~\ref{fig:hypotheses}a, requiring additional horizontal shifts for men's and women's recognition to be overlapping. 

While our findings provide strong evidence for designing intervention experiments targeting students' perceived recognition (and, in the lecture context, supplemental interventions targeting received recognition), our data do not allow us to make specific recommendations for what such interventions may look like. Future research will need to identify causal mechanisms of our correlational findings, answering the question: \textit{why} do men and women internalize their peer recognition differently? This research will likely require both in-depth qualitative methods that identify possible mechanisms (e.g., using observations and student interviews) and controlled experiments (similar to those reported in Refs.~\cite{potvin2023counternarratives,smith2020direct}) that test individual mechanisms. Here we discuss a few plausible mechanisms to motivate this future work.

One possible mechanism relates to the role of students' academic achievement: higher performing students may, on average, internalize more of their peer recognition than low performing students because high grades may help to validate recognition from others. In our study, for example, one may predict that men report higher perceptions of their peer recognition than women if men, on average, earned higher grades than women. While performance data are not available for the data reported here, a handful of prior research studies have demonstrated that men and women obtain similar grades in introductory physics courses similar to those in our data set~\cite{brewe2010toward,henderson2017exploring,dew2021gendered}. Thus, we do not believe student performance fully explains our observed gender differences in perceived peer recognition. 

Alternatively, it could be that women experience social isolation in majority-men classes (which are common in physics~\cite{hirshfield2010she}), where they have few opportunities to receive and then internalize recognition from peers. Interestingly, however, most of the physics courses included in this study had comparable numbers of men and women, or a majority of women, enrolled (Table~\ref{tab:demographics}). Prior literature also suggests that, even in majority-men physics classes, women have comparable or more peer interactions than men~\cite{sundstrom2022interactions}, further undermining this explanation of our findings.

Yet another possibility is that men and women internalize recognition from their peers differently: either men over-estimate their peer recognition or women under-estimate their peer recognition, or a combination of the two. This phenomenon may be a form of the Dunning-Kruger effect~\cite{dunning2011dunning} where men and women have biased self-assessments and, correspondingly, biases in their perceptions of how their peers view them. Previous research has documented such gender biases in undergraduate science students' self-assessments. One study, for example, found that women who earn high grades in their physics class report similar levels of self-efficacy (i.e., beliefs about their capabilities to successfully carry out physics tasks) as men who earn much lower grades~\cite{kalendar2018selfefficacy}. Another study observed that men predict higher chemistry exam grades than women despite earning similar exam grades~\cite{karatjas2015role}.

A final explanation is related to our initial assumption that students' private thoughts about who they believe is a strong physics student (as measured through our survey) are an appropriate proxy for received recognition. This assumption hinges on whether and how peer recognition is \textit{outwardly enacted}, such as by students verbally acknowledging each other as good physics students in a conversation. Such outward recognition has been shown to be critical to physics students' trajectories~\cite{carlone2007understanding}. Students' private thoughts, however, may not be as strong a proxy for such outward recognition as we have assumed. %Consider, for example, the lab context, where we observe no gender bias in these private nominations, but a bias in perceived recognition. This pattern may be because students gave less outward recognition to women than men in their peer interactions, leading to the discrepancy. This explanation is partially supported by previous work indicating that, when privately nominating strong physics peers, men disproportionately over-select men with whom they interact, as compared to women with whom they interact~\cite{sundstromWhoWhat}. That is, students' distinct interactions with men and women peers are associated with different rates of received recognition. In the lecture context, where we observe a gender bias in private nominations, discrepancies in perceived recognition may have been exacerbated by the \textit{nature} of outward recognition, for example if women faced forms of negative recognition such as micro-aggressions~\cite{barthelemy2016gender}. In either of these cases, the hypothesis is that gender differences in perceived recognition reflect accurate internalization of outwardly received (rather than privately received) recognition: women had less outward (and/or positive) recognition to internalize and so they reported lower perceived recognition. 
Future work should directly test this assumption that survey nominations approximately reflect outward recognition, for example by asking students to both nominate their strong peers and to report who they believe \textit{other students} think is a strong physics student, alongside explanations of these responses. 

Additionally, we recommend for future research to investigate the generalizability of our results by analyzing similar data from students in other countries, at other types of institutions, and from additional gender identities. We also acknowledge that recognition from people other than peers, including family members and instructors, play an important role in students' physics identity development~\cite{hazari2010connecting,carlone2007understanding}. Future work examining received and perceived recognition from these other groups of people will build on our study to create a more complete picture of the role of recognition in physics students' outcomes.

The proposed lines of research will further illuminate how students internalize their recognition and why men and women internalize their recognition differently, as we observe in the large-scale analysis presented here. With these mechanisms, the research community can then design, implement, and evaluate instructional intervention experiments that are effective in mitigating gender differences in perceived peer recognition. Such interventions will help more women and students from other marginalized groups develop their physics identity and better succeed in their physics courses.

\bibliography{ActualVsPerceived.bib}

\clearpage

\section*{METHODS}

\subsection*{Data collection}

We recruited participants from different institutions by reaching out to physics instructors the research team knew were interested in (and/or actively conducting) physics education research. We emailed the instructors, described the study, and asked if they or any of their faculty colleagues at their institution would collect data in their courses. Instructors from eight different institutions (Cornell University, Harvard University, Drexel University, Rochester Institute of Technology, North Carolina State University, Auburn University, University of Texas at Austin, and North Carolina Agricultural and Technical State University) participated in the study. All eight institutions are PhD-granting. One school is a Historically Black College or University and one school is a Hispanic-Serving Institution. Four of the institutions are public, four are private, and three are land-grant institutions. Four institutions are in northeastern United States, three are in the southeastern United States, and one is in the southern United States.

In total, we collected data from 74 introductory level physics contexts (which we refer to as ``courses"), 49 instructional lecture courses and 25 instructional lab courses, across these institutions. Our measure of received recognition (further described below), however, is only robust to up to 30\% missing data because a student's number of received nominations as strong in the course depends on survey responses from other students in the class~\cite{smith2013structural}. Therefore, we only included the 15 lecture courses and 12 lab courses with at least a 70\% survey response rate (calculated as the percent of students enrolled in the course who filled out the survey, consented to participate in research, and were 18 years old or older) in our analysis. 

These response rates mean that there were still missing data in our study and prior work has demonstrated that survey non-respondents tend to be students with lower grades~\cite{nissen2018participation}. This response bias may impact our study results, for example by over-estimating gender differences between men and women. Men and women in our study, however, were likely missing data in the same systematic way because men and women typically earn similar grades in introductory physics courses~\cite{brewe2010toward,henderson2017exploring,dew2021gendered}). Previous research has shown that when groups are not missing data in systematically different ways, the missing data do not impact between-group comparisons~\cite{walsh2019quantifying}.

The final subset of courses is also not systematically different from the full set of courses in terms of course level, course content, or student population. The lecture content of the courses in our final data set included algebra-based and calculus-based courses in both mechanics and electromagnetism. One of the courses was a one-credit computational physics seminar for first-year physics majors. Most of the lab courses were lab components attached to these lecture courses (i.e., containing the same set of students). A few of the lab courses were distinct lab courses with their own course codes and course grades. The courses at Rochester Institute of Technology were taught in a Studio Physics style, which fluidly integrates lecture and lab content into one course component. We categorized these courses as lecture courses because most of the class time was spent on physics concepts, while most of the lab courses included in this study focused on developing experimental skills~\cite{Smith2021}.

The instructor of each participating course administered an online survey via Qualtrics either in the middle or at the end of the term (i.e., data were taken from distinct samples and did not include repeated measurements) because prior work has demonstrated that students develop a community among one another by about halfway through the semester~\cite{williams2019linking}. After this point, patterns of received peer recognition remain relatively stable (as in Ref.~\cite{grunspan2016}). Students' perceptions of their peer recognition may also change over time, however our study examines the relationship between received and perceived recognition at a given time and these two variables were always measured simultaneously (i.e., on the same survey). Therefore, our recognition data from courses implementing the survey at different timepoints are likely comparable.

Some instructors administered the survey as part of a required homework assignment or an extra credit assignment to be completed outside of class, while others provided class time for students to complete the survey. Individual instructors decided whether they wanted to provide an incentive to students for completing the survey, such as homework credit or an extra point added to their final course grade. These incentives were given to all students who completed the survey, including students who did not consent to the research.

We probed perceived peer recognition through a single item used in previous research on recognition~\cite{hazari2010connecting,lock2013physics,kalender2019gendered,hazari2013science,bottomley2022relationship}, which asks students their agreement with the statement, ``My [lab/lecture/course] peers see me as a physics person" (at Rochester Institute of Technology, we only asked about the ``course" more generally). Students selected one of ``Strongly disagree," ``Disagree," ``Neutral," ``Agree," and ``Strongly agree," which we converted to a 5-point ordinal scale with 1 representing ``Strongly disagree" and 5 representing ``Strongly agree."

As a proxy for received recognition, we asked students to nominate peers in each context who they believed were knowledgeable about the instructional material with the following prompt adapted from prior work~\cite{grunspan2016,salehi2019,bloodhart2020,sundstrom2022perceptions,sundstromWhoWhat}: ``Please list the students in this physics class that you think are
particularly strong in the [lab/lecture/course] material." These questions were in an open response format, where students typed each peer's name in a separate text box. Students could enter up to 15 peers' names for each prompt, though no student provided the maximum number of names. We may have missed some nominations due to recall bias (in which students forget each other's names), however students were also given access to the course roster to facilitate their remembering and spelling of peers' names. Students could nominate anyone in their course; for example, they were not restricted to naming peers in their specific lab or recitation section.  

Students occasionally misspelled peers' names and/or reported just a first or a last name. Therefore, we processed the text responses to match the names in the nominations to the list of survey respondents (we also asked students to provide their own first and last name on the survey). First, if the first name in the text response exactly matched, or required less than two character changes to match, a unique first name in the list of respondents, then we matched the name in the response to the corresponding respondent name. For names not yet matched, we repeated this process for last names. For names still not matched, if a student listed a peer's full name that exactly matched a name in the list of respondents, or the listed name required fewer than five character changes to match a name in the list of respondents, then we matched the name in the response to the corresponding respondent name. If the response could still not be matched to a name in the list of respondents, then that nomination was removed from our analysis. The text processing matched between 75\% and 100\% of reported names to a name in the corresponding list of respondents in each course. The percentage is lower in larger courses because there are fewer unique first and last names, reducing the probability of matching responses that only provide a first or last name.

With the matched nominations, we determined each student's received recognition by counting the number of times another student in the class nominated them as strong in the instructional material. Survey responses by, and thus nominations made by, students who did not consent to participate in research or who reported that they were under 18 years old were removed from our analysis.

We also asked for self-reported demographic information on the survey (see Table~\ref{tab:demographics}). For gender, we asked students ``What is your gender?" and students selected one of the following options: ``Man," ``Woman," ``Non-binary," ``Self-describe" (with an open response text box to fill in), and ``Prefer not to disclose." We included nominations made by students of any gender in our count of received recognition. We only report on students who self-identified as a man or a woman in our main analysis because there were not enough students from additional gender identities to make claims about them. There were 31 and 20 students who selected ``Non-binary," ``Self-describe," or multiple gender identities in lab and lecture, respectively. We included these students as an ``additional gender identities" category in our statistical models (see Table~\ref{tab:coefficients}) and present their preliminary data for the relationship between received and perceived peer recognition (i.e., replicating Fig.~\ref{fig:results}c in the main text) in the Supplementary Information.

\subsection*{Data analysis}

We conducted all data analysis in the statistical software R (\url{https://www.r-project.org/}). The analysis included all survey respondents who consented to participate in research, self-reported that they were 18 years or older, and responded to the perceived recognition question(s). 87\% of all students who responded to the survey were represented in our analysis. 

We first visualized the distributions of perceived and received peer recognition split by individual course and instructional context, lab and lecture (Fig.~\ref{fig:disaggregated}). Contrary to our expectations, there was no systematic variation in these distributions despite the very different course sizes, student populations, and institutions included in the data set. Therefore, we aggregated all courses into one data set per instructional context for our analysis, without any scaling to account for different class sizes or modifications to account for different course and institution features.

We conducted a preliminary analysis that mirrors the two separate threads of prior work by comparing (i) men's and women's perceived recognition and (ii) men's and women's received recognition across this larger data set. We then performed our main analysis, which directly compares students' gender, perceived recognition, and received recognition in order to test the three possible relationships suggested by prior work (Fig.~\ref{fig:hypotheses}).

\subsubsection*{Comparing men's and women's perceived recognition}

We used the two-sided Student's \textit{t}-test (or two sample \textit{t}-test)~\cite{nissen2018cohensd,burkholder2020stats} within each instructional context (lab and lecture) to determine whether, on aggregate, men and women reported comparable perceived recognition from their peers. The Student's \textit{t}-test assumes normal distributions of men's and women's perceived recognition and approximately equal variances of the two distributions, both of which are reasonable based on Fig.~\ref{fig:aggregateperceived}. The test determines whether the difference in men's and women's mean perceived recognition is large compared to the combined uncertainty in the difference. We calculated the test statistic \textit{t},
\begin{equation*}
    t = \frac{\bar{x}_{\text{men}}-\bar{x}_\text{women}}{\sigma_p \sqrt{\frac{1}{n_\text{men}} + \frac{1}{n_\text{women}}}},
\end{equation*}
where $\bar{x}$ represents mean, $n$ represents sample size, and $\sigma_p$ is the pooled standard deviation:
\begin{equation*}
    \sigma_p = \sqrt{\frac{(n_\text{men} - 1) \sigma_\text{men}^2 + (n_\text{women} - 1) \sigma_\text{women}^2}{(n_\text{men}+n_\text{women}-2)}}.
\end{equation*}
We then compared our test statistic to the Student's $t$-distribution to reject or not reject the null hypothesis that the difference in means between men and women is zero.

We report the test statistic, degrees of freedom, and \textit{p}-value of the \textit{t}-tests. We also report effect sizes using Cohen's $d$~\cite{nissen2018cohensd,burkholder2020stats}, which, for the Student's \textit{t}-test, is calculated as
\begin{equation*}
    d = \frac{\bar{x}_{\text{men}}-\bar{x}_\text{women}}{\sigma_p}.
\end{equation*}
We interpreted the effect sizes according to the following thresholds: $0 < d < 0.2$ (negligible), $0.2 < d < 0.5$ (small), $0.5 < d < 0.8$ (medium), and $d > 0.8 $ (large). 

%Though not reported in the main text, we also compared men's and women's perceived recognition within each individual course (see Supplementary Information Section I). We used means and their standard uncertainties to evaluate whether differences were statistically distinguishable from zero. We decided to use these qualitative interpretations because relying on $p$-values, especially for a large number of comparisons, can be problematic~\cite{cumming2013understanding}. Furthermore, our data include courses with very large and very small class sizes, and thus relying on $p$-values for interpretation likely tells us more about class size than meaningful effects.

\subsubsection*{Comparing men's and women's received recognition}

Previous work~\cite{grunspan2016,salehi2019,sundstrom2022perceptions,sundstromWhoWhat} has drawn on methods of social network analysis, particularly exponential random graph models (ERGMs), to determine whether a gender bias exists in students' nominations of strong peers. These models compare the numbers of nominations made to men versus women while taking into account the proportions of men and women in the class and other network features. One drawback of this approach is that each network, in our case each of the 27 individual courses, needs to be fit with a separate model. Therefore, we did not use ERGMs to compare men's and women's received recognition on aggregate. 

An alternative approach would be to perform more traditional statistical tests on the raw measures of received recognition. The distributions of received recognition, however, were highly skewed, with many students receiving zero nominations and only a few students receiving multiple nominations (Fig.~\ref{fig:aggregateactual}). Instead, we turned to a metric similar to that used in prior work~\cite{bloodhart2020} that compares men's and women's received recognition while also taking into account the gender composition of the courses.

The metric compares the proportions of total nominations made to men and women controlling for the proportions of men and women in the class. If men and women received proportional numbers of nominations, we would expect the observed proportions of total nominations made to men and women to equal the proportions of men and women in the class. Thus, we set our expected proportion of nominations to men and women as the proportions of men and women, respectively. We then subtracted the expected proportions from the observed proportions to determine the extent to which they differed. For example, consider a class of 20 students with 50\% men and 50\% women who, altogether, made 10 nominations: one to men and 9 to women. The observed proportions of nominations would be 10\% for men and 90\% for women. We then subtract the proportions of men and women in the class (50\%) from these observed nomination proportions to get a difference of --40\% and +40\% for men and women, respectively. We used 95\% confidence intervals on the observed nomination proportions to evaluate whether the differences between the observed and expected proportions were statistically distinguishable from zero. In our example class, the 95\% confidence intervals would be [--59\%, --21\%] and [21\%, 59\%] for men and women, respectively, indicating a statistically distinguishable difference in the proportions (the intervals do not overlap with zero). %We examine these proportions on aggregate in the main text and for each individual course (see Supplementary Information Section II).

\subsubsection*{Comparing perceived recognition to received recognition and gender}

For our main analysis, we performed two-level linear regression models~\cite{vandusen2019hierarchical} to determine the relationship between gender, perceived recognition, and received recognition. We fit models with perceived recognition as the dependent variable and included predictor variables (fixed effects) for gender, received recognition, the interaction between gender and received recognition, academic year, academic major, and race or ethnicity. The perceived and received recognition variables were treated as continuous and were not transformed (further discussed below). Each race or ethnicity was included as a separate, binary variable because students could identify as more than one race or ethnicity. Including additional demographic variables allowed us to control for other aspects of students' identities that may impact their perceived recognition, even if we do not have the statistical power to claim an effect size for these variables (e.g., due to some racial or ethnic groups having small sample sizes)~\cite{walsh2021omittedbias}.

We used unconditional models to determine whether to include course and/or institution as a random effect in the model to account for any systematic differences across courses (e.g., instructional style, student population, class size, prior preparation) and institutions (e.g., geographic location, student population). For each instructional context (lab and lecture), we ran two unconditional models: one with course as a random effect and no other predictor variables and one with institution as a random effect and no other predictor variables. We used the intraclass correlation coefficient (ICC)~\cite{vandusen2019hierarchical} from these models to determine which random intercept to include in the final model. The ICC with course as a random intercept was greater than the common threshold of 0.05 in both the lab (ICC = 0.057) and lecture (ICC = 0.177) contexts and the ICC with institution as a random intercept was less than 0.05 in both the lab (ICC = 0.025) and lecture (ICC = 0.043) contexts. Therefore, we only included course as a random effect.

We fit this model to each instructional context separately and checked the model diagnostics (see Supplementary Information). We then interpreted the coefficients to determine which of the three hypothesized relationships (Fig.~\ref{fig:hypotheses}) best described our data. Evidence for the first relationship would be that received recognition, but not gender, is a significant predictor of perceived recognition and the interaction term between gender and received recognition is not significant. Evidence for the second relationship would be that both gender and received recognition, but not the interaction between these two variables, are significant predictors of perceived recognition. Evidence for the third relationship would be that gender, received recognition, and the interaction between gender and received recognition are all significant predictors of perceived recognition.

\begin{figure}[hbt!]
    \centering
    \includegraphics[width=3in]{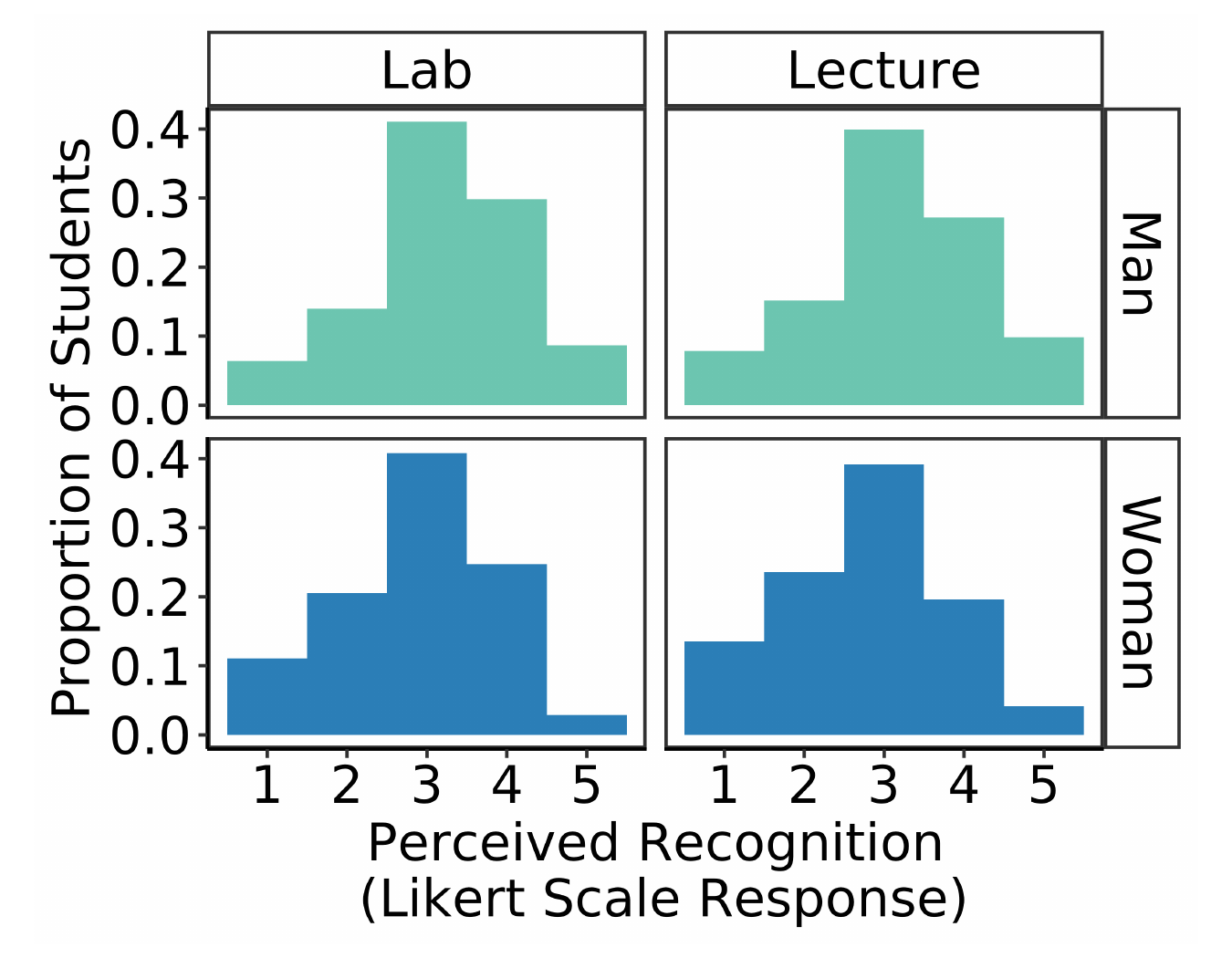}
    \caption{Histograms of perceived recognition for men and women, split by instructional context.}
    \label{fig:aggregateperceived}
\end{figure}

\begin{figure}[hbt!]
    \centering
    \includegraphics[width=3in]{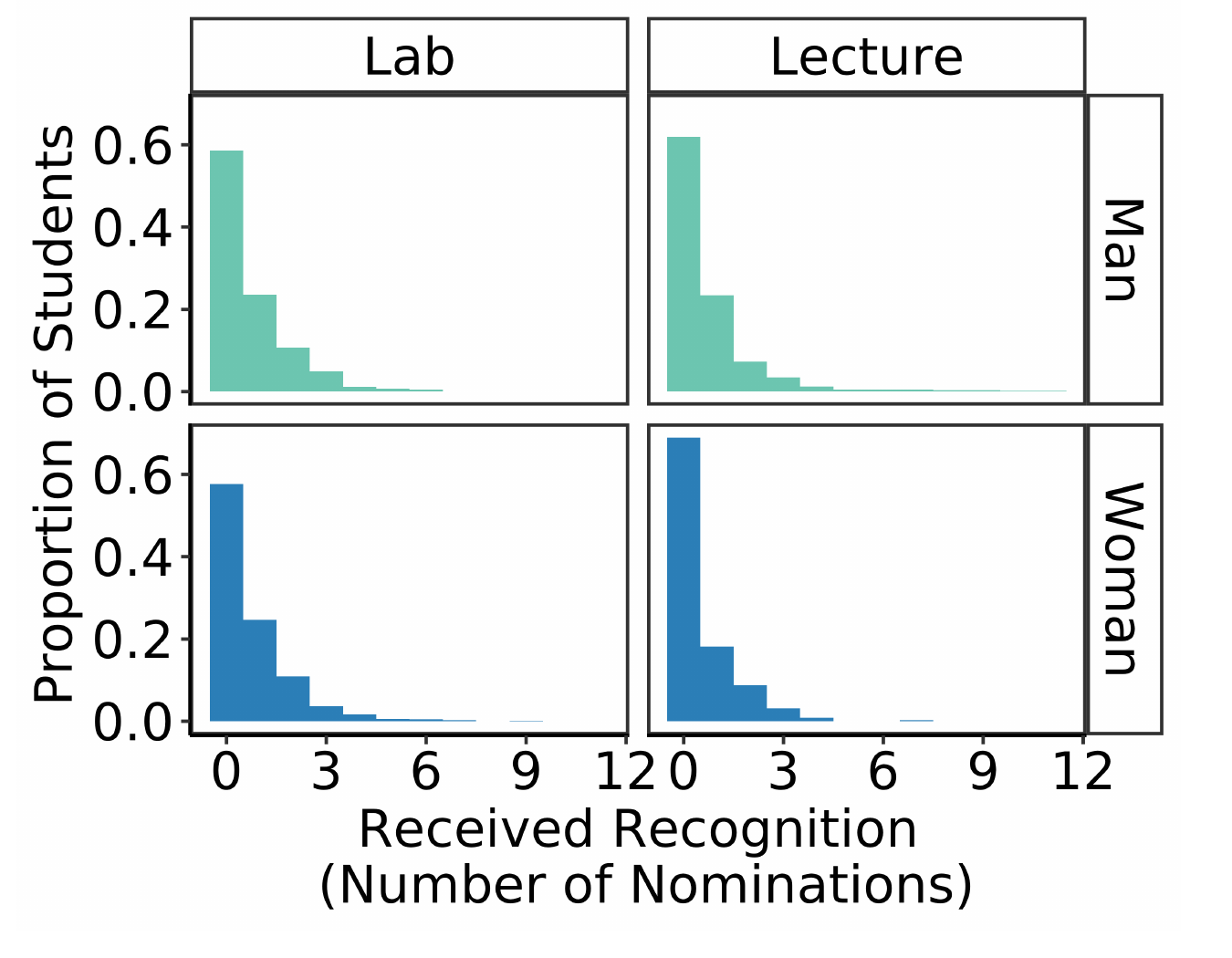}
    \caption{Histograms of received recognition for men and women, split by instructional context.}
    \label{fig:aggregateactual}
\end{figure}

We report the raw model coefficient estimates in the main text to facilitate interpretability. We do not report standardized coefficient estimates because in this particular case, both the perceived and received recognition distributions in both contexts had a standard deviation of one (Figs.~\ref{fig:aggregateperceived} and~\ref{fig:aggregateactual}). The standardized coefficients, therefore, are nearly identical to the raw coefficients--only the value of the intercept changes. This means that one can also interpret the raw coefficients as normalized coefficients: the coefficients indicate by how much a student's perceived recognition changes with \textit{either} a one unit or one standard deviation increase in the predictor variables.

We also note that the distributions of received recognition are highly skewed (Fig.~\ref{fig:aggregateactual}), with  approximately 60\% of students receiving zero nominations. This could bias the results of our linear mixed models if the data beyond zero received recognition follow a different pattern than those at zero received recognition. Thus, we checked the reliability of our results in four different ways for each instructional context: (1) running the model with all predictor variables and excluding all students receiving zero nominations, (2) running the model with just the three main predictor variables and excluding all students receiving zero nominations, (3) running the model with all predictor variables and only including a random subset of students receiving zero nominations plus all students receiving one or more nominations, and (4) running the model with just the three main predictor variables and only including a random subset of students receiving zero nominations plus all students receiving one or more nominations (see \url{https://github.com/msundstrom33/Perceived_vs_Received_Recognition}). In both lab and lecture, these four models mostly replicated the results presented in the main text: all coefficient estimates for the three main variables had the same sign and comparable magnitude to those in Fig.~\ref{fig:results}d and Table~\ref{tab:coefficients}. Some of the coefficients in these four models were not statistically significant because excluding large amounts of our data set reduces statistical power.

Finally, we note that the dependent variable--perceived recognition--could be considered ordinal rather than continuous, as it was measured on a 5-point Likert scale. Therefore, we also repeated the analysis using two-level ordinal logistic regression models, which treat the Likert scale values as ordinal~\cite{theobald2019regression}. These models obtained similar results to the linear regression models (see Supplementary Information). We report the results from the linear regression models because the coefficient estimates are more easily interpreted than logistic regression coefficients.\\

\textbf{Acknowledgements:} 
This material is based upon work supported by the National Science Foundation Graduate Research Fellowship Program Grant No. DGE-2139899 and Grant No. DUE-1836617. We thank Allison Godwin, Emily Stump, Matthew Dew, and Ashley Heim for meaningful feedback on this work. We also thank Eric Brewe, Eric Burkholder, Danny Doucette, Yasemin Kalender, Andrew Loveridge, Gregorio Ponti, and Chih Tung for their data collection efforts.

\textbf{Author contributions:} M. Sundstrom organized the data collection and conducted the data analysis. N. G. Holmes supervised the project. Both authors contributed to the development of the initial idea, the funding acquisition, and the writing of the final manuscript.

\textbf{Competing interests:} The authors declare that they have no competing interests.

\textbf{Data and code availability:} De-identified data and analysis scripts can be found here: \url{https://github.com/msundstrom33/Perceived_vs_Received_Recognition}.

\textbf{Ethics statement:} This research was approved by the Cornell University Institutional Review Board (Protocol ID\#0146224) and deemed exempt for board review as research within commonly accepted educational settings and involving educational surveys (Exemption Categories 1 and 2 of the United States Common Rule for Human Subjects Research). Informed consent was obtained from all human research participants.

%policy: https://www.nature.com/nature-portfolio/editorial-policies/ethics-and-biosecurity#Research-with-human-participants-their-data-or-biological-material

\begin{table*}[b]
\caption{\label{tab:demographics}%
Summary of all students included in our analysis. Race or ethnicity categories were not considered mutually exclusive, thus they do not necessarily sum to the total number of students. 
}
\begin{ruledtabular}
\setlength{\extrarowheight}{1pt}
\begin{tabular}{lcc}
\textrm{}&
\textrm{Lecture}&
\textrm{Lab}\\
\colrule
All  & 1059 & 1683 \\
Gender \\
\hspace{5mm}Men  & 559 & 794\\
\hspace{5mm}Women   & 480 & 858\\
\hspace{5mm}Additional (Non-binary/Self-describe/Multiple gender identities)   & 20 & 31\\
 Race or ethnicity\\
 \hspace{5mm}American Indian or Alaska Native & 10 & 19\\
\hspace{5mm}Asian or Asian American & 276 & 461\\
\hspace{5mm}Black or African American & 52 & 237\\
\hspace{5mm}Hispanic or Latinx & 126 & 239\\
\hspace{5mm}Middle Eastern or North African & 20 & 30\\
\hspace{5mm}Native Hawaiian or other Pacific Islander & 3 & 6\\
\hspace{5mm}White & 672 & 833\\
\hspace{5mm}Unknown & 20 & 38\\
Major\\
\hspace{5mm}Physics or Engineering Physics &  104 & 125\\
\hspace{5mm}Other physical science & 82 & 115
\\
\hspace{5mm}Engineering & 431 & 616
\\
\hspace{5mm}Life sciences & 251 & 515
\\
\hspace{5mm}Other & 150 & 238\\
\hspace{5mm}Unknown & 41 & 74\\
Academic year \\
\hspace{5mm}First & 408 & 486 \\
\hspace{5mm}Second & 335 & 479 \\
\hspace{5mm}Third & 248 & 553\\
\hspace{5mm}Fourth & 53 & 141\\
\hspace{5mm}Other & 9 & 14\\
\hspace{5mm}Unknown & 6 & 10
\end{tabular}
\end{ruledtabular}
\end{table*}

    \begin{table*}[hbt]
\caption{\label{tab:coefficients}%
Linear mixed models of perceived recognition with course as a random intercept for each context. Standard errors of the coefficients are shown in parentheses. The reference group for gender is men, the reference group for major is engineering, and the reference group for year is first year. Asterisks indicate statistical significance ($^{*} p < $0.05; $^{**} p < $0.01; $^{***} p < $0.001).
}
\begin{ruledtabular}
\setlength{\extrarowheight}{1pt}
\begin{tabular}{lll}
\textrm{}&\textrm{Lab}&\textrm{Lecture}
\\
\colrule
  Intercept & 3.218$^{***}$ (0.096) & 3.193$^{***}$ (0.146)\\ 
 Gender (women) & --0.237$^{***}$ (0.060) & --0.302$^{***}$ (0.075) \\
 Gender (additional gender identities) & --0.377 (0.210) & --0.378 (0.257) \\ 
 Received recognition & 0.074$^{*}$ (0.033)  & 0.108$^{***}$ (0.031) \\ 
   Gender (women) $\times$ Received recognition &  --0.011 (0.044)&  --0.012 (0.060)\\ 
   Gender (additional gender identities) $\times$ Received recognition &  --0.078 (0.204)&  0.059 (0.246)\\
   Physics major &  0.463$^{***}$ (0.097) &  0.524$^{***}$ (0.140)\\ 
      Other physical science major & --0.096 (0.102) & --0.123 (0.126)\\ 
 Life sciences or biology major  & --0.155 (0.083) & --0.183 (0.123)\\ 
 Other major & --0.161 (0.090)& --0.183 (0.124)\\ 
 Unknown major  & --0.038 (0.126) & --0.142 (0.177)\\ 
 Second year  & --0.225$^{**}$ (0.068) & --0.245$^{**}$ (0.086)\\ 
 Third year  & --0.122 (0.087) & --0.078 (0.125)\\ 
   Fourth year &  --0.236$^{*}$ (0.117)&  --0.230 (0.173) \\ 
 Other year & --0.342 (0.270) & --0.057 (0.348) \\ 
 Unknown year & --0.449 (0.316) & --0.138 (0.421) \\  
  American Indian or Alaska Native & 0.204 (0.228)  & --0.168 (0.322)\\ 
 Asian or Asian American & 0.067 (0.084) & --0.0002 (0.108)\\ 
  Black or African American & 0.029 (0.098) & --0.059 (0.159)\\
 Hispanic or Latinx  & --0.191$^{*}$ (0.084) & --0.282$^{*}$ (0.114) \\ 
  Middle Eastern or North African & --0.118 (0.184) & --0.450 (0.231)\\ 
 Native Hawaiian or other Pacific Islander & --0.467 (0.401) & 0.896 (0.581) \\ 
 White & 0.150 (0.077) & 0.132 (0.102) \\
\end{tabular}
\end{ruledtabular}
\end{table*}

\end{document}